\documentclass[prl,showpacs,twocolumn,superscriptaddress]{revtex4}
\usepackage{amssymb}
\usepackage{graphicx}

\begin{document}

\title{Comment on "Distribution of Partial Neutron Widths for Nuclei Close
to a Maximum of the Neutron Strength Function"}
\author{P.~E.~Koehler}
\thanks{corresponding author}
\affiliation{Physics Division, Oak Ridge National Laboratory, Oak Ridge, TN 37831, USA}
\author{F.~Be\v{c}v\'{a}\v{r}}
\affiliation{Charles University, Faculty of Mathematics and Physics, 180 00 Prague 8,
Czech Republic}
\author{M. Krti\v{c}ka}
\affiliation{Charles University, Faculty of Mathematics and Physics, 180 00 Prague 8,
Czech Republic}
\author{J.~A.~Harvey}
\affiliation{Physics Division, Oak Ridge National Laboratory, Oak Ridge, TN 37831, USA}
\author{K.~H.~Guber}
\affiliation{Nuclear Science and Technology Division, Oak Ridge National Laboratory, Oak
Ridge, TN 37831, USA}
\date{\today }
\maketitle

A recent Letter \cite{We2010} attempted to reconcile the disagreement
between neutron resonance data \cite{Ko2010} and random matrix theory (RMT) 
\cite{We2009}. To this end, a new formula was derived for transforming
measured ($\Gamma _{\lambda n}$) to reduced ($\Gamma _{\lambda n}^{0}$)
neutron widths for \textit{s}-wave resonances ($\lambda =1,2,...$) in
nuclides near peaks of the \textit{s}-wave neutron strength function. In
this Comment, we show that such a rescaling would not, in general, be
expected to reconcile the type of disagreement observed, and demonstrate
that indeed it does not for the specific cases in question. Hence, the
disagreements between RMT and these data remain.

The rescaling, $\Gamma _{\lambda n}^{0}=\Gamma _{\lambda n}/f^{2}(E_{\lambda
n})$, where $E_{\lambda n}$ is the resonance energy, is supposed to remove
the secular variation in $\Gamma _{\lambda n}$ with energy. The standard
rescaling, $f^{2}(E_{\lambda n})=\sqrt{E_{\lambda n}}$ was used in Ref. \cite%
{Ko2010}. The new transformation derived in Ref. \cite{We2010}, $%
f^{2}(E_{\lambda n})=\frac{1}{\pi }\left( \frac{2m}{\hbar }\right) ^{1/2}%
\frac{\sqrt{E_{\lambda n}}}{E_{\lambda n}+|E_{0}|}$, has an extra factor $%
(E_{\lambda n}+|E_{0}|)^{-1}$ arising from the single-particle state (at
energy $E_{0}$ relative to threshold) responsible for the peak in the 
\textit{s}-wave neutron strength function.

RMT predicts that \textit{s}-wave reduced neutron widths follow a
Porter-Thomas distribution (PTD) \cite{Po56}. The PTD is a $\chi ^{2}$
distribution with one degree of freedom ($\nu =1$). In Ref. \cite{Ko2010},
the maximum-likelihood (ML) method was used to obtain $\nu
_{ML}=0.57_{-0.15}^{+0.16}$, $0.47_{-0.18}^{+0.19}$, and $%
0.60_{-0.26}^{+0.28}$, for $^{192,194,196}$Pt respectively. Furthermore, it
was shown that taken together the $^{192,194}$Pt data reject the validity of
the PTD with at least 99.997\% confidence.

\begin{figure}[t]
\includegraphics[clip,width=0.80%
\columnwidth]{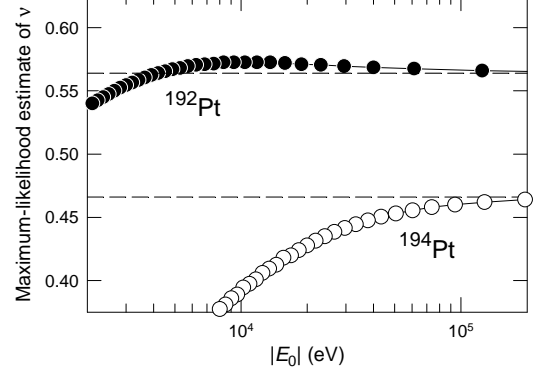} \vspace*{-0.1cm}
\caption{ML estimates of $\protect\nu $ from the sets of 153 and 161 widths $%
\Gamma _{\protect\lambda n}^{0}$ for $^{192}$Pt and $^{194}$Pt,
respectively, as functions of $|E_{0}|$. Dashed lines represent the $\protect%
\nu _{\mathrm{ML}}$ values reported in Ref.~\protect\cite{Ko2010}.}
\label{NuVsE0Fig}
\end{figure}

A smaller value of $\nu $ corresponds to a broader $\chi ^{2}$ distribution;
hence, the $^{192,194,196}$Pt data are broader than the PTD. It is easy to
see that the extra factor proposed in Ref. \cite{We2010} will, in general,
result in a broader distribution compared to the standard transformation,
except in the special circumstance when the average reduced width
(calculated using the standard transformation) is (at least approximately)
proportional to  $(E_{\lambda n}+|E_{0}|)^{-1}$. Therefore, the
transformation proposed in Ref. \cite{We2010} would not, in general, be
expected to reconcile the data of Ref. \cite{Ko2010} with the PTD, but
instead increase the disagreement. Nevertheless, we have repeated the ML
analysis using the rescaling relation derived in Ref. \cite{We2010}. The
results are shown in Fig. \ref{NuVsE0Fig}, from which it can be seen that
the new rescaling cannot reconcile the Pt data with the PTD.

According to Ref. \cite{We2010}, if the condition that $|E_{0}|$ is much
larger than the mean resonance spacing ($D_{0}=$23, 50, and 153 eV for $%
^{192,194,196}$Pt resonances, respectively) fails, it is not justified to
consider reduced neutron widths as energy-independent constants, and \textit{%
R}-matrix theory cannot be used. \ As far as we know, the necessary
alternative multi-level, multi-channel theory has not been developed, so we
cannot address this \ scenario. However, it seems very unlikely that $E_{0}$
could be this close to threshold for all three Pt isotopes.

This work was supported by the U.S. Department of Energy under Contract No.
DEAC05-00OR22725 with UT-Battelle, LLC, and by Czech Research Plans
MSM-021620859 and INGO-LA08015.

\end{document}